## Constrained Consensus Sequence Algorithm for DNA Archiving

*Dominique Lavenier*
*Univ. Rennes, CNRS-IRISA, Inria*
*dominique.lavenier@irisa.fr*

**Abstract**

The paper describes an algorithm to compute a consensus sequence from a set of DNA sequences of approximatively identical length generated by 3rd sequencing generation technologies. Its purpose targets DNA storage and is guided by specific features that cannot be exhibited from biological data such as the exact length of the consensus sequences, the precise location of known patterns, the kmer composition, etc.

# 1. Introduction

Third generation sequencing technologies provide long reads with a relatively high error rate [1]. An important task in genome assembly, for instance, is to correct these reads using the redundancy resulting from sequencing [2]. Several reads from the same part of the genome are then aligned together in order to generate a DNA consensus sequence. In principle, for de-novo genome assembly, the size and the content of the sequence are unknown.

Here we're posing the problem in a different frame. The sequence to be restored do not come from living organisms, but from sequences that have been synthesized. The target application is the DNA storage or the DNA archiving [3] [4]. Digital files are transcribed into DNA sequences with well-defined structures. For example, a file can be cut into pieces of the same size, each piece having an identifier allowing the original text to be to retrieved. The corresponding DNA sequences are therefore of known size and are surrounded by tags that are also known. In addition, since third-generation sequencing technologies are more prone to producing errors in homopolymer zones, the synthesized sequences avoid this type of situation. Consequently, in the DNA sequences, some combinations of nucleotides are forbidden.

Knowing these different constraints (size of known sequences, combination of forbidden nucleotides, known tags surrounding the sequences), the problem of building a consensus sequence can be approached differently.

# 2. Method

The input is a set of DNA sequences after base calling. We suppose that a pre-processing step has been performed to group the sequences representing the same information. The sequences have thus roughly the same size. Let $N$ be the number of sequences from which we want to find a consensus.

The main idea to get the consensus is to split the $N$ sequences into kmers and to find the best suite of overlapping kmers that represent the information to restore. The method proceeds in two main steps:

1. Construct a kmer overlap graph
2. Search constrained paths into this graph

### 2.1 Kmer overlap graph

A Kmer overlap graph (KOG) is a directed graph. Nodes of the graph are solid kmers extracted from the ***N*** sequences. A solid kmer is defined as a kmer that has been seen more than ***T*** times in a data set. Thus, the ***N*** sequences are split into overlapping kmers of size ***K***. The value of ***T***, the solid kmer threshold, depends of the number of available sequences.

An edge between two nodes n_i and n_j exists if a suffix of node n_i of size at least ***L*** is equal to the prefix of same size of node n_j. Edges are weighted by ***K-l***, ***l*** being the overlap length between n_i and n_j. The following figure exemplifies the concept.

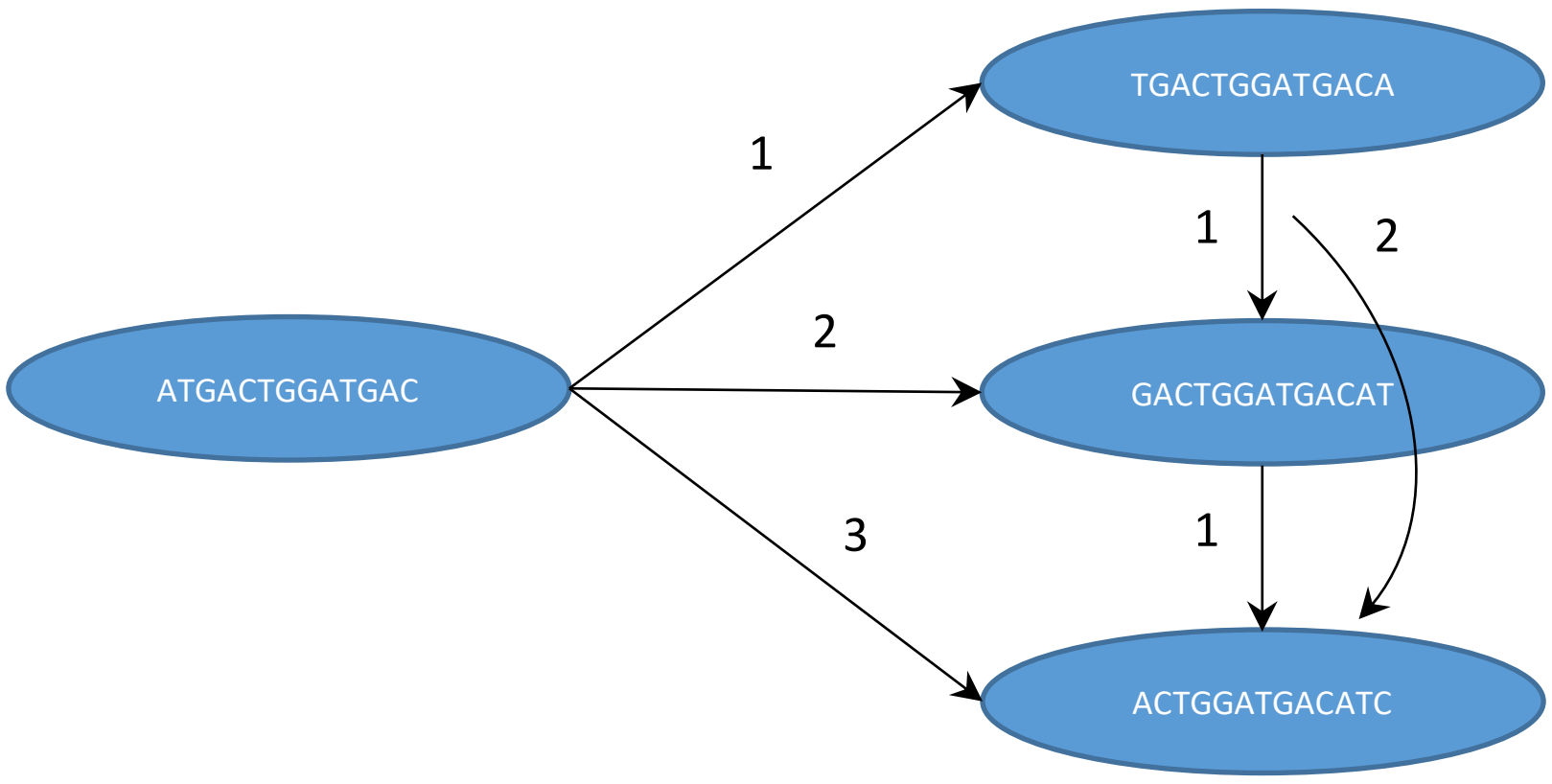

***Figure 1***: *Principle of the kmer overlap graph. K = 13. The suffix of the left node of size 12 (TGACTGGATGAC) is equal to the prefix of the upper node. The weight between these two nodes is thus equal to 13 – 12 = 1*

The expected layout of a kmer overlap graph is an elongated graph. Its shape strongly depends of the kmer size. Short values of ***K*** will tend to create loops. Large one may split the graph into different components. Figure 2 shows 3 different kmer overlap graphs built from 100 x 1 Kbp MinION sequences with 3 different values of K (10, 16, 26).

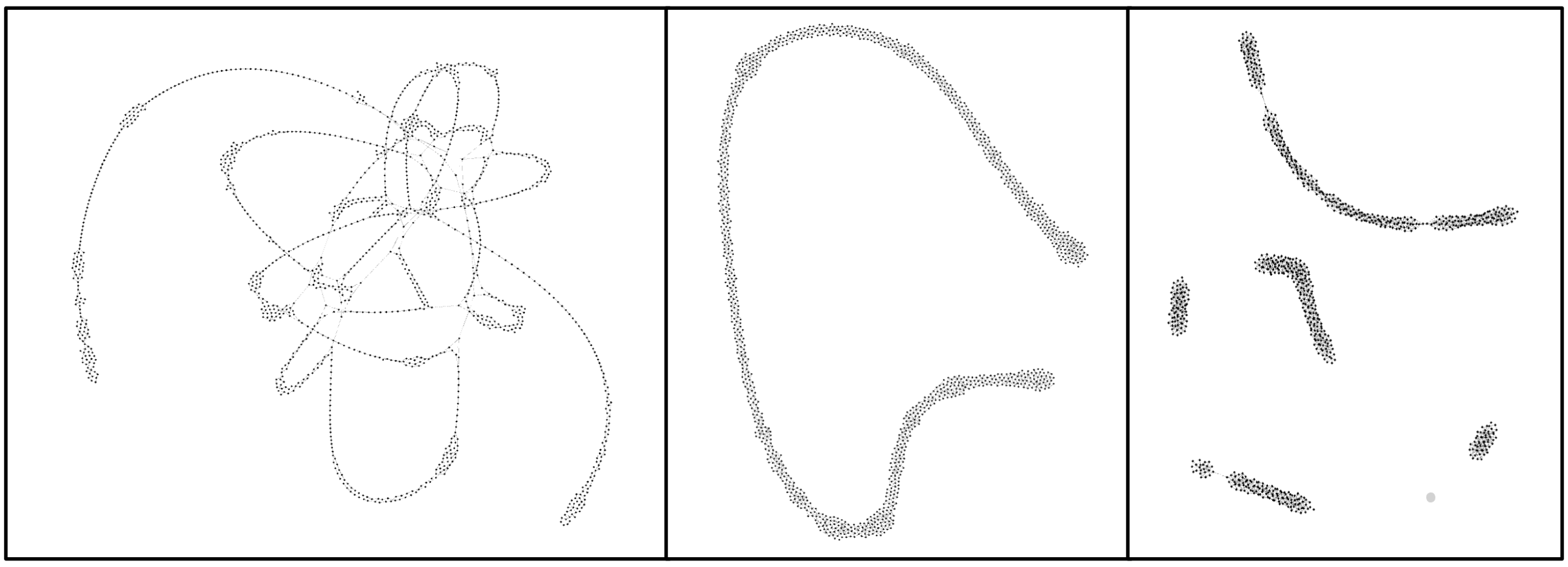

***Figure 2***: *kmer overlap graph built from 100 1Kpb MinION sequences. Left: kmer size = 10. Center: kmer size = 16. Right: kmer size = 26*

To get a graph made of a single connected component, with as lowest as possible loops, a strategy is to test different values of *K*, starting by the highest one, as follows:

```
function makeKOG (Reads)
  K = Kmax
  loop
    Skmer = extractSolidKmer(Reads,K)
    Graph = buildKmerOverlapGraph(Skmer)
    if numberOfConnectComponents(Graph) == 1 then break
    K = K-1
  return Graph
```

The number of connected components is expected to decrease with *K*. The value of *L*, the minimum overlapping size between kmers, can also influence the graph connectivity. By default, based on different experimentations, this value is set to **2*K*/3**.

To minimize loops in the graph, solid kmers have been enhanced with positions. Identical kmers that appear at different positions in the set of reads will be considered as different kmers.

### 2.2 Search constrained paths

We know that the reads are surrounded by two known short strings. We also know exactly the size of the resulting consensus sequence. As the graph contain kmers of the two surrounding strings, we chose a starting and an ending kmer. A starting kmer (`START_KMER`) is a solid kmer that belongs to the left surrounding string. Identically, an ending kmer (`END_KMER`) is a solid kmer that belongs to the right surrounding string. Having identified these two kmers, finding a consensus sequence of length ***P*** consists in finding a path of size ***P*** between `START_KMER` and `END_KMER` in the kmer overlap graph.

Paths are searched as follows: for each node of the graph, we associate a list of paths of different lengths. This list is constructed from the predecessor nodes. Suppose that we want to build the list associated to the right blue dark node (`GTGAC`) in the example below, and that the lists of the predecessor nodes have already been computed:

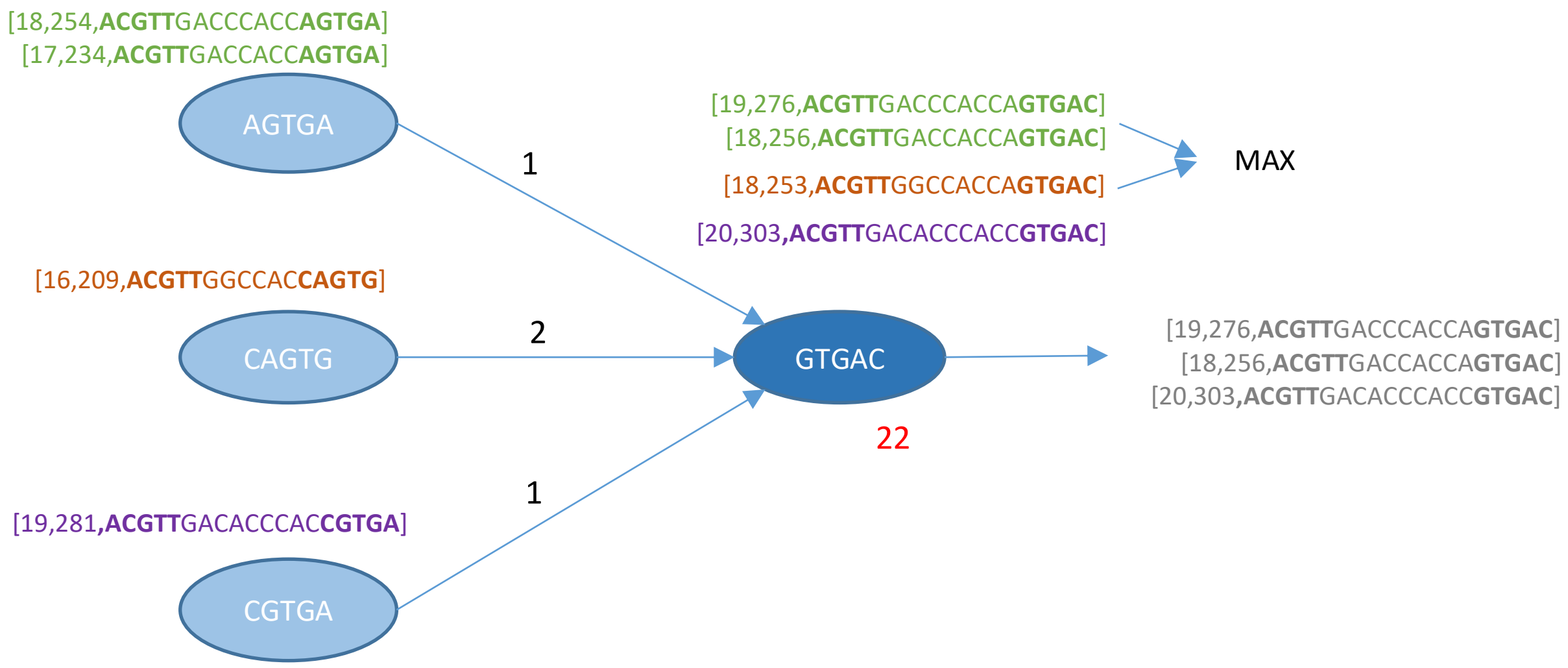

The node GTGAC has three predecessors. Node AGTGA has a list of two paths of length 18 and 17. Node CAGTG has a list of one path of length 16. Node CGTGA has a list of one path of length 19. Paths are scored according to the solidity of the kmers. The weight of a node is the number of times the kmer is present. The score of a path is the sum of the node weights from the starting kmer to the current node.

Mode precisely, to compute a list of paths, all paths from the predecessor nodes are updated according to the edge weight and the node weight of the current node. If the edge weight is $l$, the path is increased by $l$ nucleotides and the score is increased by $l$ times the node weight. If two or more paths with identical length are present, we only keep the path with the highest score.

On the above example, if we suppose that node GTGAC has a weight of 22, the path of length 17 (score = 234) of node AGTGA generates a new path of length 18 with a score of 256 (234 + 1 x 22). The path of length 16 (score = 209) of node CAGTG generates a new path of length 18 with a score of 253 (209 + 2 x 22). Thus, only the path of length 18 with the score of 256 will be kept.

All the paths begin with the starting kmer. After the search process, the END_KMER node will gather all the paths we are interested in.

**2.3 Implementation**

The method has been implemented in python and available through a software called CCSA (constrained consensus sequence algorithm). The following arguments are required:

- Reads in fastq format
- Left and right surrounding sequences in fasta format
- Length of the expected consensus sequence

The CCSA software generates a list of consensus sequences ordered by length in fasta format.

Before the construction of the kmer overlap graph, a few preprocessing steps are performed:

- Reads shorter than the expected consensus sequence length are suppressed.
- Reads are trimmed based on the surrounding sequence similarity. A read can be trimmed if, for each extremity, it has at least one kmer belonging to the left and right surrounding sequences. Reads that don't respect this rule are discarded.

The CCSA software (python code) can be downloaded here from the author web page [5]

## 3. Experimentation

The evaluation of the CCSA algorithm have been performed on several data sets of synthetic data with and without small homopolymers (<= 3 nt), and sequenced with a MinION device. The various datasets have the following properties:

| | Homopolymers | Seq length | Surrounding. Seq. length |
|---|---|---|---|
| **A_900_1** | yes | 900 | 42 / 42 |
| **A_900_2** | yes | 900 | 42 / 45 |
| **A_900_3** | yes | 900 | 41 / 42 |
| **U1_400_1** | no | 416 | 42 / 42 |
| **U1_400_2** | no | 416 | 42 / 42 |
| **U2_900_1** | no | 916 | 42 / 41 |
| **U2_900_2** | no | 916 | 41 / 42 |
| **U2_900_3** | no | 916 | 41 / 42 |

The CCSA algorithm has then been evaluated for different coverage values. For a coverage C, a subset of C reads (of length larger than the expected length) is randomly extracted from the corresponding dataset, and the CCSA software is run. The process is repeated 100 times.

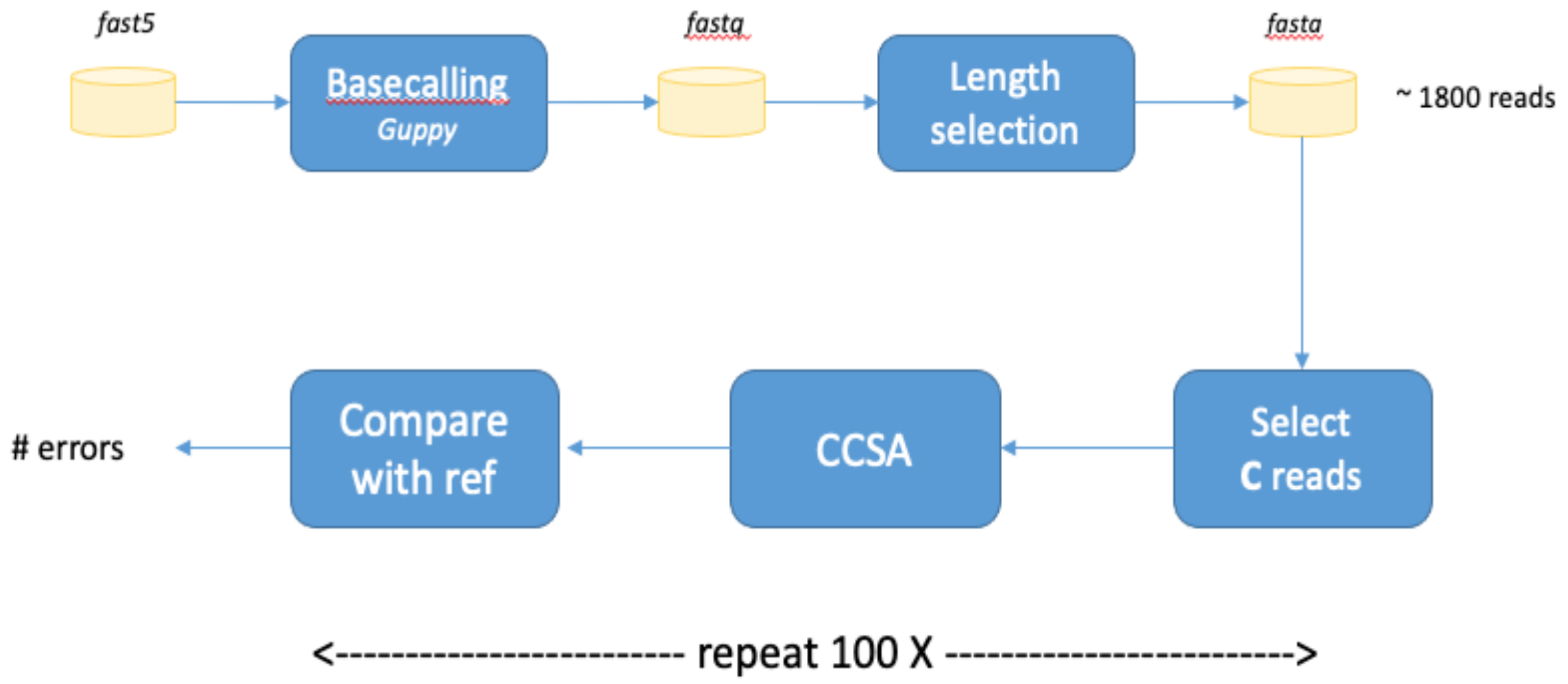

The following tables report the number of times a consensus sequence of expected length has been found (column #seq), the total number of errors (column #err), and the percentage error computed as (#seq x expected length) / #err (column %err).

**Dataset A_900: sequences with homopolymers**

| Coverage | A_900_1 | | | A_900_2 | | | A_900_3 | | |
|---|---|---|---|---|---|---|---|---|---|
| | #seq | #err | %err | #seq | #err | %err | #seq | #err | %err |
| **100** | 100 | 0 | 0 | 100 | 63 | 7.0e-4 | 100 | 34 | 3.8e-4 |
| **80** | 100 | 1 | 1.1e-5 | 98 | 46 | 5.2e-4 | 100 | 28 | 3.1e-4 |
| **60** | 100 | 2 | 2.2e-5 | 98 | 56 | 6.3e-4 | 99 | 34 | 3.8e-4 |
| **40** | 100 | 24 | 2.7e-4 | 97 | 68 | 7.8e-4 | 96 | 46 | 5.3e-4 |
| **30** | 98 | 35 | 4.0e-4 | 90 | 74 | 9.1e-4 | 87 | 46 | 6.1e-4 |
| **20** | 84 | 48 | 6.3e-4 | 75 | 65 | 9.6e-4 | 61 | 40 | 7.3e-4 |

**Dataset U_400: sequences without homopolymers**

| Coverage | U1_400_1 | | | U1_400_2 | | |
|---|---|---|---|---|---|---|
| | #seq | #err | %err | #seq | #err | %err |
| **100** | 100 | 0 | 0 | 91 | 1 | 2.6e-5 |
| **80** | 100 | 1 | 2.4e-5 | 83 | 1 | 2.9e-5 |
| **60** | 98 | 1 | 2.5e-5 | 85 | 4 | 1.1e-4 |
| **40** | 96 | 6 | 1.8e-4 | 79 | 16 | 4.9e-4 |
| **30** | 93 | 30 | 7.8e-4 | 73 | 20 | 6.6e-4 |
| **20** | 69 | 28 | 9.8e-4 | 58 | 25 | 1.0e-3 |

**Dataset U2_900: sequences without homopolymers**

| Coverage | U2_900_1 | | | U2_900_2 | | | U2_900_3 | | |
|---|---|---|---|---|---|---|---|---|---|
| | #seq | #err | %err | #seq | #err | %err | #seq | #err | %err |
| **100** | 100 | 0 | 0 | 100 | 2 | 2.1e-5 | 100 | 14 | 1.5e-4 |
| **80** | 100 | 2 | 2.1e-5 | 100 | 2 | 2.1e-5 | 99 | 12 | 1.3e-4 |
| **60** | 100 | 3 | 3.2e-5 | 100 | 5 | 5.4e-5 | 99 | 31 | 3.4e-4 |
| **40** | 96 | 111 | 1.2e-3 | 97 | 51 | 5.7e-4 | 96 | 135 | 1.5e-3 |
| **30** | 89 | 354 | 3.3e-3 | 89 | 209 | 2.5e-3 | 78 | 414 | 5.7e-3 |
| **20** | 77 | 522 | 7.2e-3 | 74 | 294 | 4.3e-3 | 60 | 638 | 1.1e-2 |

The sequences with small homopolymers provide better results. Actually, suppressing homopolymers leads to increase locally the number of short repetitions and, consequently, create small cycles in the graph that prevent correct paths to be found.

For a reasonable coverage, the error rate is quite low, but not null. Error correcting code are still needed to recover the correct information.

# Annex

## A_900_1

### Primers sequence

```
left  CCAACAACAAGGCACTCATTAACTAAGGTGGAAGCAACTGTT
right ACAACCAGTCGCTGGACAATACAGAAGTAGAAACAACTCAAC
```

### Reference sequence

```
CTCCGTCGACAGACATACCATGCCATGAACAGATACCATTCATGCCGTAAGACCGTAGTTCATCATCCGCTGCGGACCTGCTA
AGACGACGACATACCTGTATACAAACAGGACAAGACGACGTATATATAAGACCATCTATCGCACTAGTTCTATAGACCATACG
GGACCTCGTCATCCGTAACATACCGTACTGGACAGATGGACTACAATTCGACGGACATATGAACAGGACAGACAGGGACGTCA
GTCATGTAGGCAAGACGACAAACATGCGGATATCAGGCAAGATATCGAACAAGCAGAATGTATAGACATGTATAAGGCGAGTT
CGACATAAGGTTCGCGCCTGCTGAATGGATAGATAAGGCCATACGAACATGTATACAGTGCGAACCTGCAAACGTAACAGACT
GCAAGACTCATATAAGACAAACGGACCTGCTATATAAACGTAGTGAATGCCGACGTCAGATATATATACCGTGAGCTAGACAT
ACCGTACGAACCGACAAACGGGACAGGACTCGAACCGGGGCTATGAATAAGTGTAGACATGCCGTTCTAGACATATAGACCAT
CATAGGACATCATGCTAGAACGGATATACAAACCGAGACAGTGTACGGGCCGACAAACACTGAATGCCGACAGGACAAACCGT
AGATAGTGCAAGACGTACGAATAGATGAACGCCGAACGTCAGTTCTCAGGCCGACACTAAAACAAGACCGACAGGGACGAACT
CAGTGCGAGCCTCACTGAGCCATGAATAAGACTAGTTCGAGCGCCGAATATGTATGAACGCAAACATCCGATTCGGACCTCAT
CTACAAACATAGTGCAAGGCAATTCGAACTCCTTCATTCAGACAGTCATCATACCGTGAATCAGACACTA
```

## A_900_2

### Primers sequence

```
left  CCAACAACAAGGCACTCATTACAGCTGGTAATGATGGAACAT
right TTCTTTCAGATTTTTCAAGTTGGACAGAAGTAGAAACAACTCAAC
```

### Reference sequence

```
AGGGATAAGAACAGTCGCCACTGGGCAGTCATTCGAACCAATCCGCGCAACACAGGGCAACATAAGGCGTAACGCATGCTTCC
ATACCTCATATCATGACAAAGTCTGTGCAGGACCGGACCGCTCGACAGACCTCGGATCGCAAACCGGGATGGACTACAGTGGA
TAGTCATCCTCAGGACCGAATAGAATGCTAGCGCACACTGAGCTCATTCGCCTGCGGGATACGGATCGTAGGGCATGGAACGT
AAACGGATGTGAACAGACATCATTCGACGGGACAAACGTCTAGTAAACGTACCAGTCGCCGAGCGTATCTAGCTGCAACAAAC
GTACCGTGCAAGTGCCTCAACGAATGTGACTAAGAATAAGGCTATACCGTGCAAGCATATAGGACCAGTATGAGACTCCGCGC
CATACAATCCGGGGACTGGGACACTGTAACCTAAGATATGGAATCATCCGAACACCGGGCCTCAGGCGCGCGTCAAGTCGACC
TACAAGGCCTTCATATCAAGTCAACAGGCGACAAACAGGGCAGTGCGGATCAAGGGCCTCAGGCAAAACAGAATACCGAACGA
TTCAAGGCAAGTTCAAATAAGTGGGCGTAACGGATAGTATGACGTATGATTCTCCTGGGCGGTGAGACAGTCTACTGAGCCTA
CTAGAACGCGAATACAGACACTATACTAGTAGACTAAGAACATATCAGACAAGGCCGACTCCGTACGCGAATGTCGCCGTAGA
CGCGCGAATGTACGCGAACAGTCGCAGATATCCGTAGAGTTCAACAGGGCCTCGACCGGTGCCTCGACGTGTATACCGGTCGC
GCAGACGAATCAGTCAGTGACATCCTCGCCGACACCGACAGGACCGGGATATGGACGCGACGTGCTAGA
```

## A_900_3

### Primers sequence

```
left  CCAACAACAAGGCACTCATTACAATTCCATCTGTCTTCATTC
right ACTCAACACAATTAACAACGGACAGAAGTAGAAACAACTCAAC
```

### Reference sequence

```
GACGAACATCAGACCATCTCAGGTTCATTCATGTCAGACTCAGATATATGAGTTCGACAGACATTCGTATACCATCAGACAGG
TTCATCGAACTCATCATATGAGCAGACATATAAGTTCATGCCATCCGTCAGGACGTAGACGAACGTGCCATCAGACTCGAATA
GTGTATACCATCGCGGACTCATGTAGGTTCGTAGACATTCGGTTCATCAAACGAATAGACATGGACCATCGAACAGGCAGACT
CATCAGACGGACCGACAAGACAGGACAGACGGACGAACATCAGGACAAGACATCGAACAGACATCAAACATATGCCATCTCAT
CATATAGACCATTCGACAGGACCGACATATAGGACGAACATAGACATATATGAACTCGAACCATCGCCATGCCGTGTAGACGT
ATAGTCGAATAGTCAAACATATGTAGACGAACGTGAATATAGACCGTAGTCGTCAAACGAACCGTATAGGATACCATCAGACT
CGAACGCCGTGTAGTGTAGGATATAGGACATCATCCGTCGTCATGTCATCGAACGTGAACGTAGGTTCGTGGACTCGACAGGA
TACCATGTCATCATATAGTCAGTTCGTCTAGACTCGAACAGGTTCGTCAGACTCAGGTTCGGTTCGTCAGATATAGTGTATAG
GACATTCAGGACCATCAGGTTCATTCATCATATCAGACGTTCATCGACATCAGGACAGGCAAACATAGTGGACTCGCGTGAAC
AAGACATGAGCATCAGGCGAACTCAAACAGACGGACGAACATGCCATCATCAGACGTTCGTATAGGATATAGTCAGGCAGACA
GGATAGGACAAACGAACAGACATGTATATGAGCCATCGAACCATCAGACAAACGAACGTCAGACCGTAGT
```

## U1_400_1 / U1_400_2

### Primers sequence

```
left  TAGTGTCATGTATCTCGTCTCATCGTCGTATCTCGTCATCTA
right GCGCTCGACAGATACAGACATCGTGAGCGACGAGCGACGAGC
```

### Reference sequence (U1_400_1) - noHP

```
GCGACTCATCTGCAGCTCGAGTACAGTCTCTGTAGCGTGTACGATCACATGCATACATGATGTGCATGAGCATGTGTACACAT
ACGATAGTGAGATGTGACTATCGACGAGTCACAGATACTCTCTCAGATAGTACGCACGAGTCAGCAGTCAGAGCGACGCGCAG
CTCACATCGCGTATGTCTGATATATGTGCGAGATGCTCTCACGACAGAGATACACATCAGTCGACTACAGATACACGTATACA
GTAGTAGCAGTCTGAGAGATCGACTGCAGCACTGCTCGTACGTCGTGTGTATGCACGTGAGCTGTGCATCTACGATGATGAGT
GTATGTATATCGATCAGTGTAGCGCATGTAGTCATCGATACTACGCACGATGACACACTAGTCAGTACATCACTCATGTCGTG
T
```

### Reference sequence (U1_400_2) - DiffMax

```
GCGACTCGTCTGCTGATCTAGTATAGTCTCTGTCGCGTCGACGATCTCATGTATGTATCGTGTGCATCAGTATGTGTACTCGT
GTGATAGTGTGATGTGAGTGTCGATGTGTCACAGATGATCTCTCAGATCGTATGTACTCGTCTGCAGTCTGTGCGTCTCGCAT
GTCACGTCTCGTGTGTCTGATAGATGTGCTCGATGATCTCATCTCAGTGATAGACGTCAGTCGACTATCGATAGACGTCGACA
GTAGTATCAGTCTGTCAGATCGACTATCTCAGTGATCGTATGTCGTGTGTATGTACGTCTGCAGTGTATCGACGATGATGTGT
GTATCGATATCGATCAGTGTAGTGTATCTAGTCGTCGATAGTATGTACGATGTCACTCTAGTCAGTGTATCTCTCGTGTCGTC
T
```

## U2_900_1

### Primers sequence

```
left  CCAACAACAAGGCACTCATTAGTTTATATTTGACTGGCGAAG
right CTGACTCTGCGATCAAGACACAGAAGTAGAAACAACTCAAC
```

### Reference sequence (U2_900_1) - noHP

```
GCGACTCATCTGCAGCTCGAGTACAGTCTCTGTAGCGTGTACGATCACATGCATACATGATGTGCATGAGCATGTGTACACAT
ACGATAGTGAGATGTGACTATCGACGAGTCACAGATACTCTCTCAGATAGTACGCACGAGTCAGCAGTCAGAGCGACGCGCAG
CTCACATCGCGTATGTCTGATATATGTGCGAGATGCTCTCACGACAGAGATACACATCAGTCGACTACAGATACACGTATACA
GTAGTAGCAGTCTGAGAGATCGACTGCAGCACTGCTCGTACGTCGTGTGTATGCACGTGAGCTGTGCATCTACGATGATGAGT
GTATGTATATCGATCAGTGTAGCGCATGTAGTCATCGATACTACGCACGATGACACACTAGTCAGTACATCACTCATGTCGTG
TATCTGTGACGAGTCTCGATCGCGCTGTCTGATGCAGTCGAGACGACGTACACAGTACGCAGCTCGCTCGTGCTGTGACTCAT
GCAGAGCAGAGCGCTAGAGACTAGAGCACGTAGTGCTGTCTATCGACTCGTCACATATCATCTGTCACTCTGTAGTGACAGTG
ACTACTCTCGTAGCTAGTGACGCTACAGCGCAGCTCGCTGAGCTACGTGAGTATCACATAGACTCAGTCTATCTACTGTCGAC
TCGTGATCACAGTCTCGTGAGTGTGTCTAGCTGAGTCTGATGACGTGTACGTCTCACACTGTCACTCAGCATATACGAGCAGC
ACTATATGCGACTCGTAGCGTCGCACGCTCACAGATGCGCAGTCACTATCTCTACAGCTCACATACACGTCTGTGCTCTGCTA
TCACTAGAGACTCTCGCACTCGTAGCGCTGATGTCTGTATCTACGCACTATCATATATGCATGATAGTCTCTCAGCGTGCTGT
CAG
```

## U2_900_2

### Primers sequence

```
left  CCAACAACAAGGCACTCATTAGCAGCGAAAGCCAACTTAGA
right GCATACGACGTCACAAGGAAACAGAAGTAGAAACAACTCAAC
```

### Reference sequence (U2_900_2) - incHPMax

```
GGTTACATGTCGCGTCGACTAGATCGTGTCTGTCTGTCTGTGTCTAGTGCAGTATCTCGTGCTCTAGATGTATAGATAGATAG
ACGATGTGTGCTCAGTGTGTGATCTCTCTAGATGAGTATGAGTGTAGTGTATCGATCTCATAGATACAGTCGTACAGATCATG
ATGTACTGCGTGTCTGTCTGTCATAGAGCACTCGTCGATGATATACTGATCGTGACGATGTGCGCGCGTGTGTCATGTCTGTA
GTCTGTCTGTCTGTGCGCTCGATATGTCGCACTGTGCAGACAGATGCTGATGATAGAGTCGTGCTATCTCATACAGACAGATA
GCACGTGTAGCTCGAGCGCTCACTCACTCTCGTGCAGCAGTAGTACTCGTCTCTAGTGATAGATAGATAGATATGTGACGTAG
TATAGTGTGCGTATCGACATCGTGCGATAGACTCGACGTGTCTGACTATCTGTCTCAGCGATAGTGTAGAGTCACTCACAGCT
AGTACTCAGTGACTGATACGAGACGCGCTGAGCGATATGCGCGTGTCTCTAGATCGTACGTGCGTAGTGTGTAGTCGTGCGTC
TCTGTCTGTGCTCGATCGTCAGTGAGTCTCGACACGCTGAGCGATATCTCTCAGTGTAGTGTAGATGTCTCGCTAGATAGATA
GTGTCTGATGTCGTGATGTGCTCGTGATAGACACTCACTCTACTCGACATGCTACGCTAGAGTCGTCATGCTCACTAGTCTCG
```

```
ACGTCACAGACAGACATAGAGTGTGTGAGTCACGACTATCGCGCATGCTATCTGTCTGTCGATGATCTCACTCACTCACTCTC
ATAGACAGTGAGATACTACTGATACGCTGCTATCTGTCTATACAGATCGTATATCACGTGCAGCAGTACACTCGTAGTGCTAT
CAC
```

## U2_900_3

### Primers sequence

```
left  CCAACAACAAGGCACTCATTAGAACCTGTTACAGCACAGC
right CATTTAACCCTCGTCCAGCACAGAAGTAGAAACAACTCAAC
```

### Reference sequence (U2_900_3) - noHPmax

```
GCGACTCATCTGCAGCTCGAGTATAGTCTCTGTAGCGTGTACGATCGCATGCATGCATGATGTGCATGAGCATGTGTACACAT
GCGATAGTGAGATGTGACTATCGACGAGTCGCAGATACTCTCTCAGATAGTACGCACGAGTCAGCAGTCAGAGCGACGCGCAG
CTGATATCGCGTATGTCTGATATATGTGCGAGATGCTCTCACGATATAGATACACGTCAGTCGACTACAGATACACGTCTACA
GTAGTAGCAGTCTGAGAGATCGACTGCAGCAGTGCTCGTACGTCGTGTGTATGCACGTGAGCTGTGCATCTACGATGATGAGT
GTATGTATATCGATCAGTGTAGCGTATGTAGTCATCGATAGTACGCACGATGACACACTAGTCAGTACATCACTCGTGTCGTG
TATCTGTGACGAGTCTCGATCTCGCTGTCTGATGTAGTCGAGACGACGTACACAGTACGCAGCTCGCTCGTGCTGTGACTGAT
GTAGAGCAGAGCTCTAGAGACTAGAGCGCGTAGTGCTGTCTATCGACTCGTCACATGTCATCTGTCACTCAGTAGTGACAGTG
ACTACTCTCGTAGCTCGTGACGCTACAGCGTAGCTCGCTGAGCTACGTGAGTATCACATCGACTCAGTCTATCTACTGTCGAC
TCGTGATCGCAGTCTCGTGAGTGTGTCTAGCTGAGTCAGATGACGTGTACGTGTCACAGTGTCACTCAGCATATACGAGCAGC
AGTATATGCGAGTCGTAGCGTCGCACGCTCACAGATGCGCAGTCACTATCTCTACAGCTGATATACACGTGTGTGCTCTGCTA
TCACTAGAGAGTGTCGCACTCGTAGCGCTGATGTCTGTATCTACGCGCTATCATATATGCATGATAGTCTCTCAGCGTGCTGT
CAG
```